\documentclass[draftclsnofoot,onecolumn]{IEEEtran}
\usepackage{bm}

\usepackage{cite}

\ifCLASSINFOpdf
   \usepackage[pdftex]{graphicx}
\else
   \usepackage[dvips]{graphicx}
\fi
\usepackage{subfigure}

\usepackage{amsmath}
\newtheorem{theorem}{\bf Theorem}
\newtheorem{lemma}{\bf Lemma}

\usepackage{amssymb}

\usepackage{algorithmic}
\usepackage{array}
\ifCLASSOPTIONcompsoc
  \usepackage[caption=false,font=normalsize,labelfont=sf,textfont=sf]{subfig}
\else
  \usepackage[caption=false,font=footnotesize]{subfig}
\fi
\usepackage{caption2}

\usepackage{stfloats}
\usepackage{url}
\begin{document}
\title{Energy Efficiency Optimization for \\UAV-assisted Backscatter Communications}

\author{Shengzhi Yang, 
        Yansha Deng*, 
        Xuanxuan Tang, 
        Yuan Ding, 
        and~Jianming Zhou % <-this % stops a space
\thanks{Manuscript received June 26, 2019; accepted July 25, 2019 by IEEE Communication Letter. The associate editor coordinating the review of this paper and approving it for publication was D. Ciuonzo. The work of Shengzhi Yang was supported by the China Scholarship Council.  \textit{(Corresponding author: Yansha Deng.)}}
\thanks{Shengzhi Yang and Jianming Zhou are with School of Information and Electronics, Beijing Institute of Technology, Beijing 100081, China. (E-mail: shengzhi.yang@kcl.ac.uk; E-mail:zhoujm@bit.edu.cn)}% <-this % stops a space
\thanks{Yansha Deng is with King's College London, London WC2B 4BG, UK. (E-mail: yansha.deng@kcl.ac.uk)}
\thanks{Xuanxuan Tang is with Air Force Early Warning Academy, Wuhan 430019, China. (E-mail: tang\_xx@126.com)}
\thanks{Yuan Ding is with Heriot-Watt University, Edinburgh EH14 4AS, UK. (E-mail: yuan.ding@hw.au.ck)}% % <-this % stops a space
}


\markboth{}%
{Shell \MakeLowercase{\textit{et al.}}: Bare Demo of IEEEtran.cls for IEEE Journals}



\maketitle

\begin{abstract}
Future Internet-of-Things (IoT) has high demand for energy-saving communications, especially in remote areas and smart cities. To meet this demand, we propose novel Unmanned Aerial Vehicle-assisted backscatter communications, where a UAV first collects data from multiple terrestrial backscattering tags via time division multiple access, and then flies into the coverage region of a terrestrial base station to upload its collected data to its associated base station. To determine the optimal UAV data collection location, we first analyze the system average outage probability, and then optimize the energy efficiency with the optimal backscattering location through Golden Section method under UAV energy constraint. Our analytical and simulation results illustrate that there is a trade-off between UAV data collection location and the outage probability, and the optimal UAV data collection location to achieve maximum energy efficiency needs to be closer to the tags for lower UAV transmit power.
\end{abstract}


% \fi
\IEEEpeerreviewmaketitle

\section{Introduction}

Backscatter communication is a promising technology for future IoT networks to link a huge number of smart devices in various applications, including industrial automation, precision agriculture, and smart cities \cite{7876867,8170328}. In ambient backscatter systems \cite{Kellogg:2014:WBI:2619239.2626319}, ambient radio frequency (RF) energy, such as TV, WiFi and cellular signals, is harvested as the only power source for tag operations. Since the available ambient energy is limited, its communication range is commonly in the range of a few meters, hindering its extensive field applications \cite{LoRa_Backscatter}. In order to extend its communication range, bistatic architecture with dedicated RF power sources \cite{7327131,LoRa_Backscatter,Lorea_back} is proposed, where a nearby signal generator is exploited to create a RF carrier that, after being modulated by the tags, is able to convey information to readers located hundreds or even thousands of meters away. Specifically, in \cite{LoRa_Backscatter}, their experimental results illustrated when the backscattering device is close to an RF source, the system transmission distance can reach 2.8km. 
%The authors in \cite{7876867} analyzed the success probability and the transmission capacity of a large-scale device-to-device backscatter communication network using stochastic geometry.

Thanks to the high maneuverability, ease of deployment, hovering ability and low cost, Unmanned Aerial Vehicles (UAVs) are more attractive to provide wireless connectivity \cite{7463007}, especially for applications in remote areas and restricted regions, such as intelligent agriculture in large farms and fauna and flora protection in national parks \cite{8470897}. In these scenarios, UAVs can act as information collectors and uploaders from IoT devices to the nearest base station (BS), while powering the IoT nodes simultaneously. A UAV-based prototype wireless power transfer (WPT)-BS in \cite{7953846} demonstrated the ability of UAVs both as the WPT and communication platforms. Besides, the UAV relay location \cite{8068199} and the UAV power consumption \cite{8329973,8743402} for energy efficiency communications can be optimized. 

Motivated by above, the mobility of the UAV can be optimized compared with backscattering communications at fixed locations to investigate the optimal system performance. We propose a UAV-assisted backscatter communications, which combines the advantages of both UAVs and backscattering communications. A UAV acts as a data collector from multiple terrestrial backscattering tags via time division multiple access (TDMA). Then the UAV deposits the collected data into a far-away BS after a period of flight. We first analyze the outage probability of this UAV-assisted backscatter communication, and then optimize the UAV data collection location for maximum energy efficiency under UAV energy constraint. Our results show that lower UAV transmit power leads to closer UAV data collection location to the tags so as to decrease the outage probability and improve the energy efficiency.

The rest of the paper is organized as below. Section \ref{section_systemmodel} describes the system model, and then provides analysis of the system average outage probability. The UAV data collection location is optimized to achieve maximum energy efficiency in Section \ref{section_optimization}, and validated via simulation in Section \ref{section_results}. Conclusions are drawn in Section \ref{section_conclusion}.

\textit{Notation}: A modified incomplete gamma function is denoted by $\Gamma (m,x) = \frac{1}{\Gamma(m)}\int\limits_0^x {{t^{m - 1}} \cdot {e^{ - t}}\mathrm{d}t}$. $\Gamma (x)$ denotes the Gamma function. $f_\textsc{a}(x)$ and $F_\textsc{a}(x)$ denotes a probability density function (PDF) of A, and the corresponding cumulative distribution function (CDF), respectively. 

\section{System Model and Outage Probability}
\label{section_systemmodel}

\subsection{System Model}

\begin{figure*}[!htb]
\centering
\includegraphics[width=4.5in]{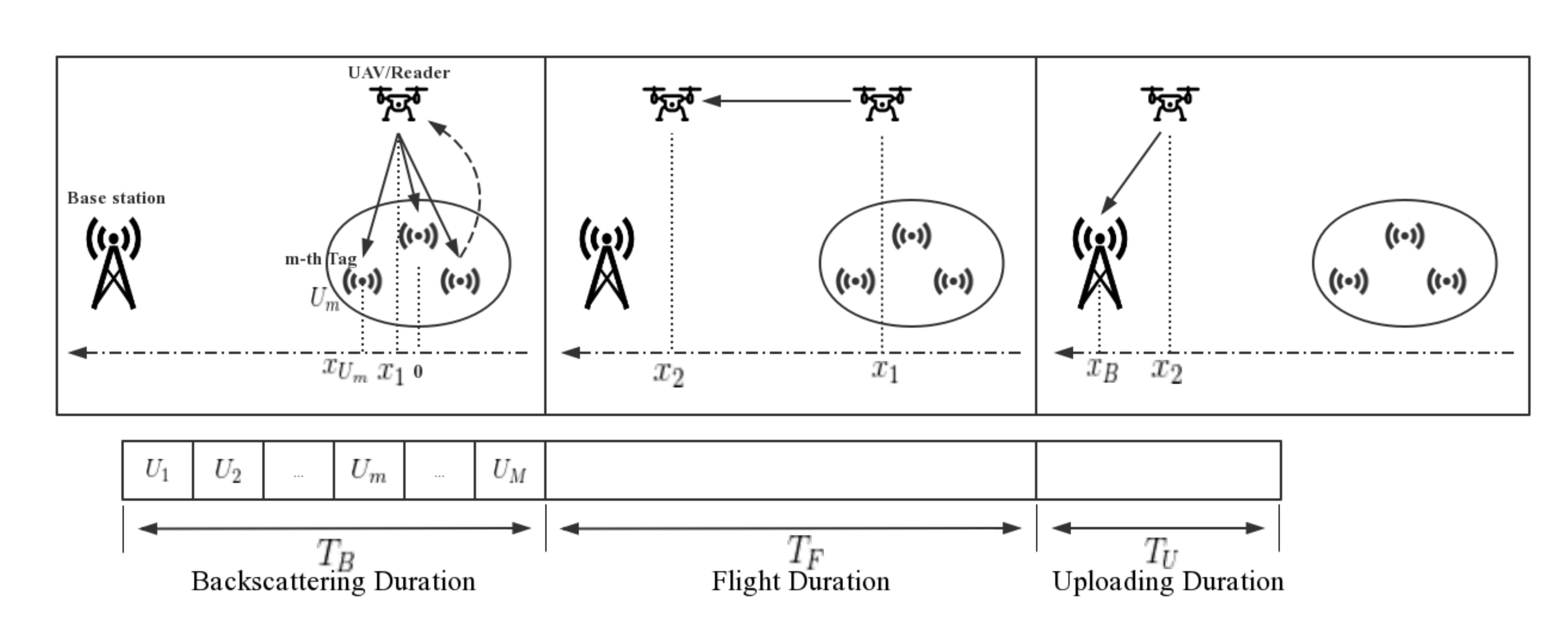}
\centering
\caption{System model}
\label{SysModel}
\end{figure*}
As illustrated in Fig. \ref{SysModel}, we consider a UAV-assisted backscatter communication system, where a UAV with the limited total energy $E_\textup{total}$ collects data from $M$ terrestrial tags. We assume that these tags ${U_m}$ ($1\le m \le M$) are randomly scattered with uniform distribution within a range of 20 meters. The entire operation consists of three stages: information collection via backscattering communications, UAV flight to nearby BS, and UAV uplink information to BS.

The backscatter communication period $T_\textsc{b}$ is equally divided into $M$ time slots, and the time sequence is given. Each terrestrial tag $U_m$ at $(x_{\textsc{u}_m},0)$ is chosen one slot to delivery their collected information to the UAV via backscattering. In the other $M-1$ time slots, the tag only harvests energy. The UAV hovers at ($x_1,h$) to collect data with its fixed transmit power $P_\textsc{v}$ from the backscattering tags via TDMA. As shown in Fig. \ref{SysModel}, the distance between the UAV data collection location and the $m$-th terrestrial tag is ${d_{\textsc{v{u}}_m}}= \sqrt {{h^2} + {{({x_1} - {x_{\textsc{u}_m}})}^2}}$ with an elevation angle ${\theta _{\textsc{v{u}}_m}}= \frac{{180^\circ}}{\pi } \times \arcsin (\frac{h}{{d_{\textsc{v{u}}_m}}})$. 

After collecting the data from all the tags, the UAV flies to the location $(x_2,h)$ near a BS at location $(x_\textsc{b},0)$ with a flight duration $T_\textsc{f}$ using a constant velocity of $v$. In the uploading duration $T_\textsc{u}$, the UAV uploads collected data to the BS. As shown in Fig. \ref{SysModel}, the distance for uploading from the UAV to the BS is $d_\textsc{vb}= \sqrt {{h^2} + {{({x_2} - {x_\textsc{b}})}^2}}$ with an elevation angle ${\theta _\textsc{vb}} = \frac{{180^\circ}}{\pi } \times \arcsin (\frac{h}{{{d_\textsc{vb}}}})$.

Let us use $\mathrm{VB}$ to represent \textup{UAV to BS} and $\mathrm{VU}_m$ to represent \textup{UAV to tag $U_m$}, and denote $a\in \{\mathrm{VB}, \mathrm{VU}_m\}$ and $b\in \{\mathrm{LoS}, \mathrm{NLoS}\}$. We assume that all the channels are Nakagami-$m$ fading channels (i.i.d) with shape factor $k_{a,b}$, which is a generalized channel model \cite{8533585}. Therefore, the PDF of the channel power gain $|g_{a,b}|^2$ follows Gamma distribution, 
\begin{equation}
	\label{cdfgain}
	f_{|g_{a,b}|^2}(x) = \frac{1}{{\Gamma ({k_{a,b}})}} \cdot {\left(\frac{{{k_{a,b}}}}{{\Omega}}\right)^{{k_{a,b}}}} \cdot {x^{^{{k_{a,b}} - 1}}}  \mathrm{e}^{ \left(\frac{{ - {k_{a,b}}}}{{\Omega}} x \right)},
\end{equation}
where $\Omega$ represents the mean value, i.e. $\Omega = \omega {\beta _0}d_a^{ - \alpha}$ if path loss is considered, where the path loss exponent $\alpha=2$ \cite{7841993}, $\omega=1$ for LoS propagation and $\omega=\eta_a$ for NLoS propagation, and $\beta_0$ denotes the channel power gain at the reference distance $d_0 = 1$m. In addition, the line-of-sight (LoS) probability \cite{6863654} is given as 
\begin{equation}
	\label{pVUm}	
	{p_{a,\textsc{LoS}}} = \frac{1}{{1 + c \cdot\exp [ - q({\theta _{a}} - c)]}}, 
\end{equation}
where $c$ and $q$ are constant values depending on the environment. Accordingly, non line-of-sight (NLoS) probability is 
\begin{equation}
	\label{pVUm_NLOS}
	{p_{a,\textsc{NLoS}}} = 1-p_{a,\textsc{LoS}}. 
\end{equation}

%For large-scale transmission, we use the free space propagation model $g_f^a = \sqrt{\omega \beta_0d_a^{-2}}$ with pass loss factor $\alpha=-2$ \cite{7841993}, where $\omega=1$ for LoS propagation and $\omega=\eta_a$ for NLoS propagation, and $\beta_0$ denotes the channel power gain at the reference distance $d_0 = 1$m. We also assume that all the small-scale fading channels follow Nakagami-$m$ distribution (i.i.d) with \textit{shape factor} $k_{a,b}$. Therefore, The probability density function (PDF) of the channel power $|g_{a,b}|^2$ is  

%Thus, the channel models for LoS and NLoS links are written as \vspace{-2mm}
%\begin{equation}
%\label{gain}
%  g_a = \sqrt{\omega \beta_0d_a^{-2}}\cdot g_{a,b},
%\end{equation}
%with LoS probability given in Eq. (\ref{pVUm}) and NLoS probability given in Eq. (\ref{pVUm_NLOS}), respectively.
We assume the UAV works in a full duplex mode\footnote{This assumption is commonly used in most researches \cite{6836141, 7913737}. With this assumption, perfect signal separation can be operated at the UAV. Thus, there is no interference at the UAV.}. Within the $m$-th time slot of the backscattering duration, considering a discrete-time signal model in the baseband, the received signal at the UAV is given by ${y_{{\textsc{u}_m}\textsc{v}}} = {g_{\textsc{v}{\textsc{u}_m}}}{{g'}_{\textsc{v}{\textsc{u}_m}}}{s_1}{s_3} + {g_{\textsc{v}{\textsc{u}_m}}}{n_{{\textsc{u}_m}}} + {n_\textsc{v}}$ \cite{6836141}, where ${s_1}$ is the signal transmitted by the UAV, $s_3$ is the tag's information signal, ${n_\textsc{v}}$ denotes the additive white Gaussian noise (AWGN) with its power $\sigma_{\textsc{v}}^2$ at the UAV, $n_{\textsc{u}_m}$ denotes the AWGN with its power $\sigma_{\textsc{u}_m}^2$ at the tag $U_m$. Besides, the received power at the UAV is given by ${P'_\textsc{v}} = \eta_\textsc{r} P_\textsc{v} \left| g_{\textsc{vu}_m}\right|^2 \left|g'_{\textsc{vu}_m} \right|^2$. Therefore, its signal-to-noise-ratio (SNR) can be derived as 
\begin{equation}
\label{gammaUmV}
\begin{aligned}
\gamma_{\textsc{u}_m\textsc{v}} = \frac{\eta_\textsc{r} P_\textsc{v} \left| g_{\textsc{vu}_m}\right|^2 \left|g'_{\textsc{vu}_m} \right|^2} {\left| g_{\textsc{vu}_m} \right|^2 \sigma_{\textsc{u}_m}^2 + \sigma _\textsc{v}^2},
\end{aligned} 
\end{equation}
 where $\eta_\textsc{r}$ is the fraction of reflected power.

By the end of $M$ time slots, the total data collected by UAV is ${T_{\textsc{u}}}{R_{\textsc{u}}} = \sum\limits_{m = 1}^M {\frac{{{T_{\textsc{b}}}}}{M}{R_m}},$ where $R_m$ and $R_\textsc{u}$ are the information rate of the $m$-th tag and the UAV, respectively. In addition, $\gamma _{{\rm{th}}}^m \buildrel \Delta \over = {2^{{R_m}}} - 1$ and $\gamma _{{\rm{th}}}^{\textsc{u}} \buildrel \Delta \over = {2^{{R_{\textsc{u}}}}} - 1$ are the SNR thresholds corresponding to their Shannon capacity per Hz. Besides, the signal from the UAV to the BS can be characterized by ${y_\textsc{vb}} = \sqrt {P_\textsc{v}}\cdot g_\textsc{vb}\cdot s_2 + {n_\textsc{b}}$, where $s_2$ is the transmit signal from the UAV and $n_\textsc{b}$ is the additive white Gaussian noise of uploading channel with its power ${\sigma_\textsc{b}}^2$. Thus, the SNR of the \textup{UAV-BS} signal can be written as
\begin{equation}
\label{gammaVB}
{\gamma_\textsc{vb}} = \frac{P_\textsc{v}\cdot \left| g_\textsc{vb} \right|^2}{{\sigma_\textsc{b}}^2}.
\end{equation}

\subsection{Energy Outage Probability}
For the $m$-th terrestrial tag $U_m$, the total energy received from the UAV consists of the energy received during backscattering single time slot and the energy harvested during other $M-1$ time slot, which can be  given as 
\begin{equation}
\label{epsilonm}
\begin{aligned}
{\varepsilon _m} &= \frac{{\left( {1 - {\eta _{\textsc{r}}}} \right){\eta _{\textsc{c}}}{P_\textsc{v}}{{\left| {{g_{\textsc{vu}_m}}} \right|}^2}}}{M}{T_{\textsc{b}}} + \frac{{\left( {m - 1} \right){\eta _{\textsc{c}}}{P_\textsc{v}}{{\left| {{g_{\textsc{vu}_m}}} \right|}^2}}}{M}{T_{\textsc{b}}}\\
 &= \frac{{( {m - {\eta _{\textsc{r}}}} ){\eta _{\textsc{c}}}{P_\textsc{v}}{{| {{g_{\textsc{vu}_m}}}|}^2}{T_{\textsc{b}}}}}{M},
\end{aligned}		
\end{equation}
 where $\eta_\textsc{r}$ is the fraction of reflected power in the backscattering period, and $\eta_\textsc{c}$ is the circuit conversion efficiency\cite{7876867,8170328}. The energy outage events occur if the energy received by the $m$-th terrestrial tag from the UAV $\varepsilon _m$ in Eq. (\ref{epsilonm}) is less than the circuit power loss during the backscattering time slot, which can be formulated as ${P_{\textsc{e},\textup{m}}} = \Pr ( {{\varepsilon _m} < \frac{{{T_\textsc{b}}}}{M}{P_\textsc{c}}} )$, where $P_\textsc{c}$ is the circuit power loss at each terrestrial tag.

\begin{lemma}[Energy Outage Probability]
\label{Lemma_Penm}
The energy outage probability of the $m$-th tag can be derived as  
\begin{equation}
\label{Penm}
\begin{aligned}
&{P_{\textsc{e},\textup{m}}} = {p_{\textsc{vu}_m,\textsc{LoS}}} \cdot \Gamma \left( {{k_{\textsc{vu}_m,\textsc{LoS}}},\frac{{{P_\textsc{c}}\cdot d_{\textsc{vu}_m}^2\cdot {k_{\textsc{vu}_m,\textsc{LoS}}}}}{{(m - {\eta _\textsc{r}}){\eta _\textsc{c}}{P_\textsc{v}}{\beta _0}}}} \right)\\ 
&+ {p_{\textsc{vu}_m,\textsc{NLoS}}} \cdot \Gamma \left( {{k_{\textsc{vu}_m,\textsc{NLoS}}},\frac{{{P_\textsc{c}}\cdot d_{\textsc{vu}_m}^2\cdot {k_{\textsc{vu}_m,\textsc{NLoS}}}}}{{(m - {\eta _\textsc{r}}){\eta _\textsc{c}}{P_\textsc{v}}{\beta _0}{\eta _{\textsc{vu}_m}}}}} \right), 
\end{aligned}
\end{equation}
where  ${p_{\textsc{vu}_m,\textsc{LoS}}}$ and ${p_{\textsc{vu}_m,\textsc{NLoS}}}$ are given in Eq. (\ref{pVUm}) and Eq. (\ref{pVUm_NLOS}), respectively. 
\end{lemma}

\textit{Proof:} See Appendix. \ref{Proof_Penm}.$\hfill\blacksquare$

\subsection{System Average Outage Probability}

Note that there are two main factors causing information outage, which are the energy outage, and that the SNR of the backscattered signals at the UAV to the BS is lower than a given threshold. Thus, the outage probability between the $m$-th tag and the BS can be defined as

\begin{equation}
\label{Pinm}
\begin{aligned}
{P_\textup{in,m}} =& [1 - \Pr ({\gamma _{{\textsc{u}_m}\textsc{v}}} \ge {\gamma _\textup{th}^m},{\gamma _\textsc{vb}} \ge \gamma _\textup{th}^\textsc{u})](1 - {P_{\textsc{e},\textup{m}}})+P_{\textsc{e},\textup{m}}\\
 =& 1-\left[1-{F_{{\gamma _{{\textsc{u}_m}\textsc{v}}}}}(\gamma _\textup{th}^m)\right]\left[1-{F_{{\gamma _\textsc{vb}}}}({\gamma _\textup{th}^\textsc{u}})\right]\left[1 - {P_{\textsc{e},\textup{m}}}\right],
\end{aligned}
\end{equation}
where ${P_{\textsc{e},\textup{m}}}$ is given in Eq. (\ref{Penm}) in Lemma \ref{Lemma_Penm}. Meanwhile, the system average outage probability can be defined as  
\begin{equation}
\begin{aligned}
\label{averagePin}
P_\textup{in}=\frac{1}{M}\mathop \sum \limits_{m = 1}^M P_{\textup{in,m}}. 
\end{aligned} 
\end{equation}
Then, by substituting Eq. (\ref{Penm}), Eq. (\ref{FgammaUmV}) with $x=\gamma _\textup{th}^m$ and Eq. (\ref{FgammaVB}) with $x=\gamma _\textup{th}^\textsc{u}$ into Eq. (\ref{Pinm}), the system average outage probability is finally derived in the following theorem. 

\begin{theorem}[The System Average Outage Probability]
\label{Theo_pin}
The closed-form expression for the system average outage probability of the UAV-assisted backscattering communications in Nakagami-$m$ fading is derived as 
\begin{equation}
\begin{aligned}
\label{Pin}
&P_\textup{in}=1-\frac{1}{M}\mathop \sum \limits_{m = 1}^M [1-{p_{\textsc{vb},{\textsc{LoS}}}}\cdot \Gamma ( {{k_{\textsc{vb},{\textsc{LoS}}}},\frac{{{\gamma _\textup{th}^\textsc{u}}\sigma _\textsc{b}^2d_\textsc{vb}^2\cdot {k_{\textsc{vb},{\textsc{LoS}}}}}}{{{P_\textsc{v}}{\beta _0}}}})\\
&-{p_{\textsc{vb},{\textsc{NLoS}}}}\cdot \Gamma ( {{k_{\textsc{vb},{\textsc{NLoS}}}},\frac{{{\gamma _\textup{th}^\textsc{u}}\sigma _\textsc{b}^2d_\textsc{vb}^2\cdot {k_{\textsc{vb},{\textsc{NLoS}}}}}}{{{P_\textsc{v}}{\eta _\textsc{vb}}{\beta _0}}}} )]\\
&\times [1-{p_{\textsc{vu}_m,\textsc{LoS}}} \cdot \Gamma ( {{k_{\textsc{vu}_m,\textsc{LoS}}},\frac{{{P_\textsc{c}}\cdot d_{\textsc{vu}_m}^2\cdot {k_{\textsc{vu}_m,\textsc{LoS}}}}}{{(m - {\eta _\textsc{r}}){\eta _\textsc{c}}{P_\textsc{v}}{\beta _0}}}} )\\
&-{p_{\textsc{vu}_m,\textsc{NLoS}}} \cdot \Gamma ( {{k_{\textsc{vu}_m,\textsc{NLoS}}},\frac{{{P_\textsc{c}}\cdot d_{\textsc{vu}_m}^2\cdot {k_{\textsc{vu}_m,\textsc{NLoS}}}}}{{(m - {\eta _\textsc{r}}){\eta _\textsc{c}}{P_\textsc{v}}{\beta _0}{\eta _{\textsc{vu}_m}}}}} )]\\
&\times [1-\int_0^\infty  ({p_{\textsc{v}{\textsc{u}_m},{\textsc{LoS}}}}\cdot \Gamma ( k_{\textsc{vu}_m,\textsc{LoS}},\frac{{{\gamma _\textup{th}^m}\left( {y\sigma _{{\textsc{u}_m}}^2 + \sigma _\textsc{v}^2} \right)d_{\textsc{v}{\textsc{u}_m}}^2}}{{{\eta _{\textsc{r}}}{P_\textsc{v}}y{\beta _0}}})\\
&+{p_{\textsc{v}{\textsc{u}_m},{\textsc{NLoS}}}}\cdot \Gamma ( k_{\textsc{vu}_m,\textsc{NLoS}},\frac{{{\gamma _\textup{th}^m}\left( {y\sigma _{{\textsc{u}_m}}^2 + \sigma _\textsc{v}^2} \right)d_{\textsc{v}{\textsc{u}_m}}^2}}{{{\eta _{\textsc{R}}}{P_\textsc{v}}\cdot y\cdot {\eta _{\textsc{v}{\textsc{u}_m}}}{\beta _0}}}))\\
&\cdot({p_{\textsc{v}{\textsc{u}_m},{\textsc{LoS}}}}\cdot{{f_{{{\left| {{g_{\textsc{v}{\textsc{u}_m}}},\textsc{LoS}} \right|}^2}}}\left( y \right)}+{p_{\textsc{v}{\textsc{u}_m},{\textsc{NLoS}}}}\cdot{{f_{{{\left| {{g_{\textsc{v}{\textsc{u}_m}}},\textsc{NLoS}} \right|}^2}}}\left( y \right)})\textup{d}y],
\end{aligned} 
\end{equation}
with $p_{a,b}$ given in Eq. (\ref{pVUm}) and Eq. (\ref{pVUm_NLOS}), and $f_{|g_{a,b}|^2}(y)$ given in Eq. (\ref{cdfgain}), $a\in \{\mathrm{VB}, \mathrm{VU}_m\}$ and $b\in \{\mathrm{LoS}, \mathrm{NLoS}\}$.
\end{theorem}

\section{UAV Data Collection Location Optimization}
\label{section_optimization}

The average capacity of the proposed system is ${\frac{1}{M}\mathop \sum \limits_{m = 1}^M {R_m}(1 - {P_\textup{in,m}})}$ with total energy cost of ${T_\textsc{f}}{P_\textsc{f}} + ({T_\textsc{b}} + {T_\textsc{u}}){P_\textsc{v}}$, where $P_\textsc{f}$ is the consumed power of UAV during fly. Thus, the optimization problem of energy efficiency under energy constraint can be formulated as 
\begin{equation}
\label{origin_opti_problem_ineq}
  \begin{aligned}
  &\max\limits_{T_\textsc{f}}\quad \eta_\textup{en}=\frac{{\frac{1}{M}\mathop \sum \limits_{m = 1}^M {R_m}(1 - {P_\textup{in,m}})}}{{{T_\textsc{f}}{P_\textsc{f}} + ({T_\textsc{b}} + {T_\textsc{u}}){P_\textsc{v}}}}\\
  &\begin{array}{r@{\quad}r@{}l@{\quad}l}
  s.t.&{T_\textsc{f}}{P_\textsc{f}} + ({T_\textsc{b}} + {T_\textsc{u}}){P_\textsc{v}} \le {E_\textup{total}}.\\
   \end{array}
  \end{aligned}
\end{equation}

The problem (\ref{origin_opti_problem_ineq}) can be rewritten as 
\begin{equation}
\label{origin_opti_problem_ineq_re}
  \begin{aligned}
  &\max\limits_{x_1}\quad \eta_\textup{en}(x_1)=\frac{{\frac{1}{M}\mathop \sum \limits_{m = 1}^{M} {R_m}(1 - {P_\textup{in,m}(x_1)})}}{{{\frac{x_2-x_1}{v}}{P_\textsc{f}} + ({T_\textsc{b}} + {T_\textsc{u}}){P_\textsc{v}}}}\\
  &\begin{array}{r@{\quad}r@{}l@{\quad}l}
  s.t.&\frac{{{x_2} - x_1}}{v}{P_\textsc{f}} + ({T_\textsc{b}} + {T_\textsc{u}}){P_\textsc{v}} \le {E_\textup{total}}.\\
   \end{array}
  \end{aligned}
\end{equation}

Due to the complexity of ${P_\textup{in,m}(x_1)}$, the first- and the second-order derivatives cannot be derived readily. Thus, one-dimensional linear searching method is utilized. We first derive the feasible region $\left\{ {{x_1}|{x_1} \ge \frac{{{E_\textup{total}} - ({T_\textsc{b}} + {T_\textsc{u}}){P_\textsc{v}}}}{{{P_\textsc{f}}}} \cdot v} \right\}_.$ Then, the search region of one-dimensional linear searching method to find the optimal solution of $\eta_\textup{en}(x_1)$ in the problem (\ref{origin_opti_problem_ineq_re}) is $\left\{ {{x_1}|{x_1} \ge \frac{{{E_\textup{total}} - ({T_\textsc{b}} + {T_\textsc{u}}){P_\textsc{v}}}}{{{P_\textsc{f}}}} \cdot v} \right\} \cap \left\{ {{x_1}| {x_1} \le {x_2}} \right\}.$ Golden Section Method is used to find the optimal solution $x_1^*$ of Problem (\ref{origin_opti_problem_ineq_re}) within the aforementioned searching region \cite{book8675}. Due to the limit of the space of the paper, we do not give the details.

\section{Simulation Results}
\label{section_results}

In this section, we validate our derived analytical expressions and conduct performance analysis based on numerical simulations with the following parameters: $M=3$; $P_\textsc{c}=0.001$W; $\eta_\textsc{r}=0.5$; $\eta_\textsc{c}=0.5$; $\eta _\textsc{vb} = 0.5$; $\eta _{\textsc{vu}_m} = 0.5$; $h=50$m; $v=10$m/s, which are common assumptions in previously reported works. Other parameters are set as follows: $T_\textsc{u}=1$s; $T_\textsc{b}=1$s; $x_2=300$; $x_\textsc{b}=500$; $c=11.95$; $q=0.136$; $\sigma_{\textsc{u}_m}^2=\sigma_{\textsc{v}}^2=10^{-9}$W; ${P_\textsc{f}} = 100$W; $R_m = 1$; $\beta_0 = 1$. The shape-factor $k_{a,b}$ is 2.  
%\begin{table}[!htb]
%\renewcommand{\arraystretch}{1.3}
%\caption{System Parameters}
%\label{Tab_Parameters}
%\centering
%\begin{tabular}{|l|c|}
%\hline
%Parameters & Setting \\
%\hline
%\hline
%UAV transmit power & $P_\textsc{v}=$40dBm, 37dBm, 30dBm\\
%\hline
%Shape factor & $k=2$\\
%\hline
%Number of terrestrial tags & $M=3$\\
%\hline
%UAV upload location & $x_2=300$m\\
%\hline
%Bases station location & $x_\textsc{b}=500$m\\
%\hline
%Circuit power loss & $P_\textsc{c}=0.001$W\\
%\hline
%Fraction of reflected power & $\eta_\textsc{r}=0.5$\\
%\hline
%Energy conversion efficiency & $\eta_\textsc{c}=0.5$\\
%\hline
%Uploading duration & $T_\textsc{u}=1$s\\
%\hline
%Backscattering duration & $T_\textsc{b}=1$s\\
%\hline
%UAV flight height & $h=50$m\\
%\hline
%Parameters for LoS probability & $c=11.95$ and $b=0.136$\\
%\hline
%UAV flight power & $P_\textsc{f}=50$W\\
%\hline
%UAV flight speed & $v=10$m/s\\
%\hline
%The noise variance & $\sigma^2=10^{-9}$W\\
%\hline
%The iteration times of Monte-Carlo & $10^5$\\
%\hline
%\end{tabular}
%\end{table}

\begin{figure}[!htb]
\centering
\includegraphics[width=2.5in]{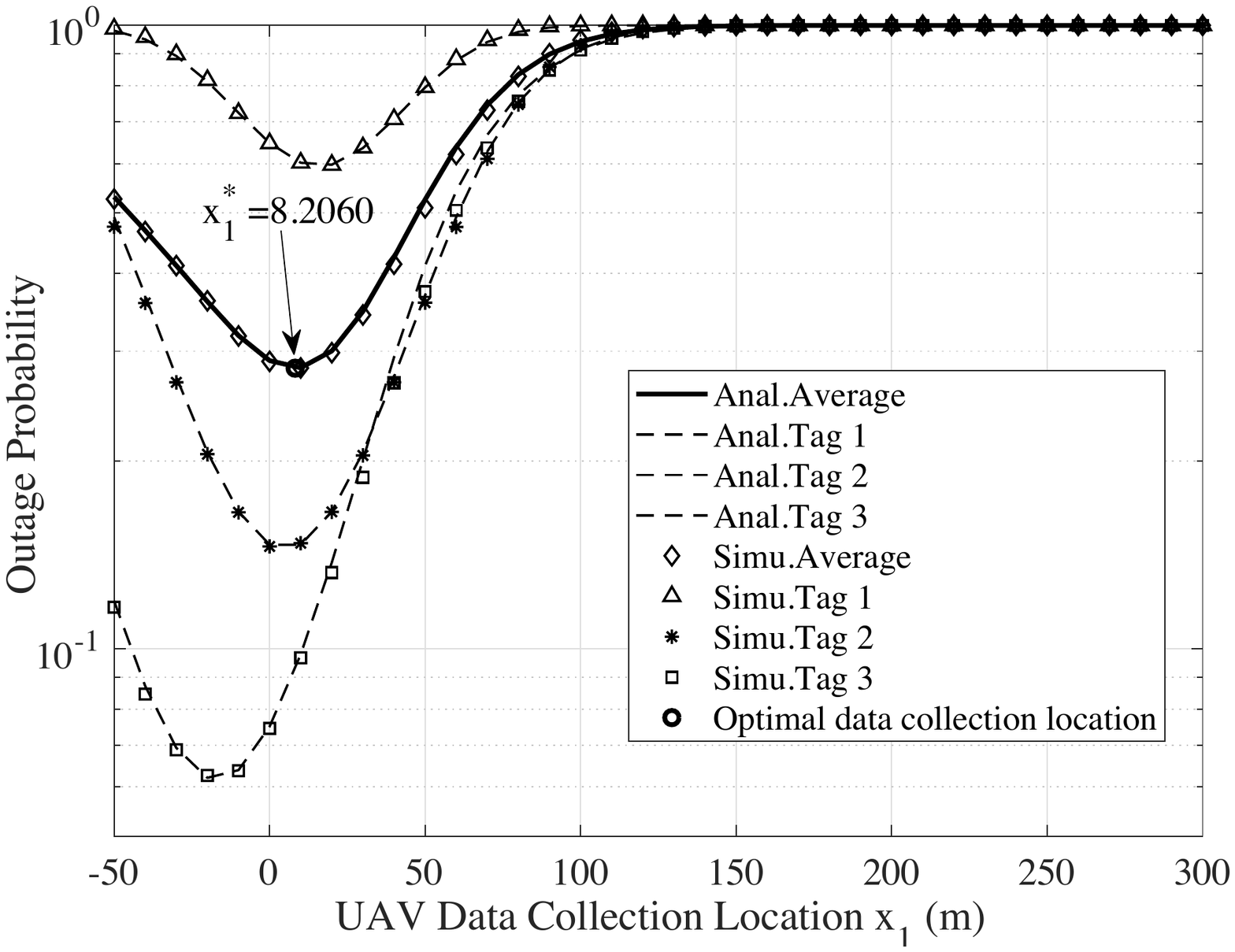}
\centering
\caption{System average outage probability $P_\textup{in}$ v.s. UAV data collection location $x_1$.}
\label{Fig_InforOutPro}
\end{figure}
Fig. \ref{Fig_InforOutPro} plots the impact of UAV data collection location on the system average outage probability $P_\textup{in}$ given in Eq. (\ref{Pin}) and the outage probability of each tag $P_\textup{in,m}$ given in Eq. (\ref{Pinm}). We see that the theoretical results match well with the Monte-Carlo simulations. It is obvious that the problem in Eq. (\ref{Pin}) is convex, and there exists an optimal collection location leading to a minimum outage probability. The terrestrial tag with larger backscattering distance to the UAV suffers higher outage probability.

%\begin{figure}[!htb]
%\centering
%\includegraphics[width=3in]{2.eps}
%\centering
%\caption{Optimal UAV data collection location $x_1^*$ under the energy constraint with different UAV transmit power.}
%\label{EEwithInCons}
%\end{figure}
\begin{figure}[!h]
\centering
\subfigure[Energy efficiency v.s. UAV data collection location $x_1$ under the energy constraint with different UAV transmit power$P_\textsc{v}$.]{
\includegraphics[width=2.5in]{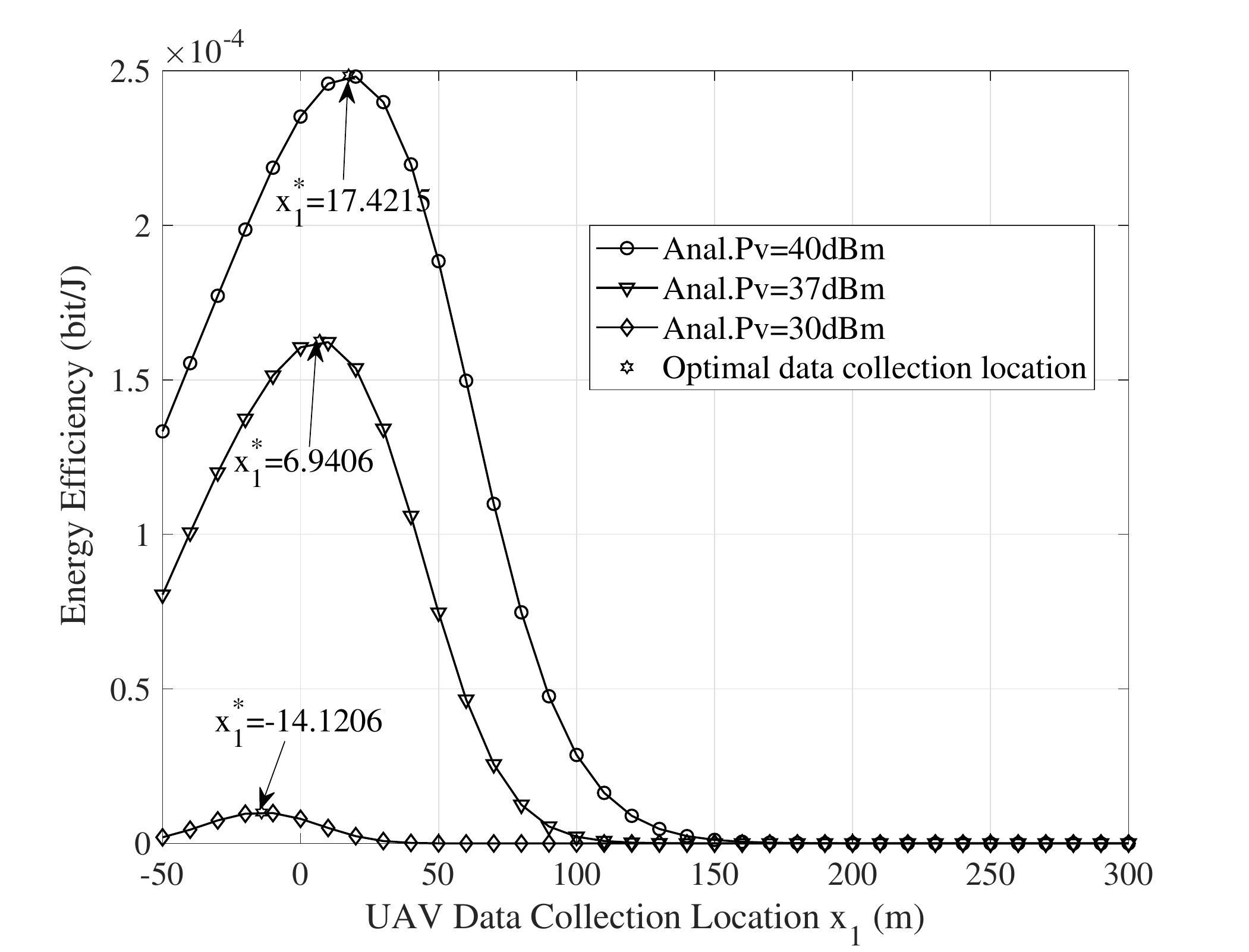}}
\subfigure[Energy efficiency v.s. UAV transmit power $P_\textsc{v}$ under the energy constraint with different UAV data collection location $x_1^*$.]{
\includegraphics[width=2.5in]{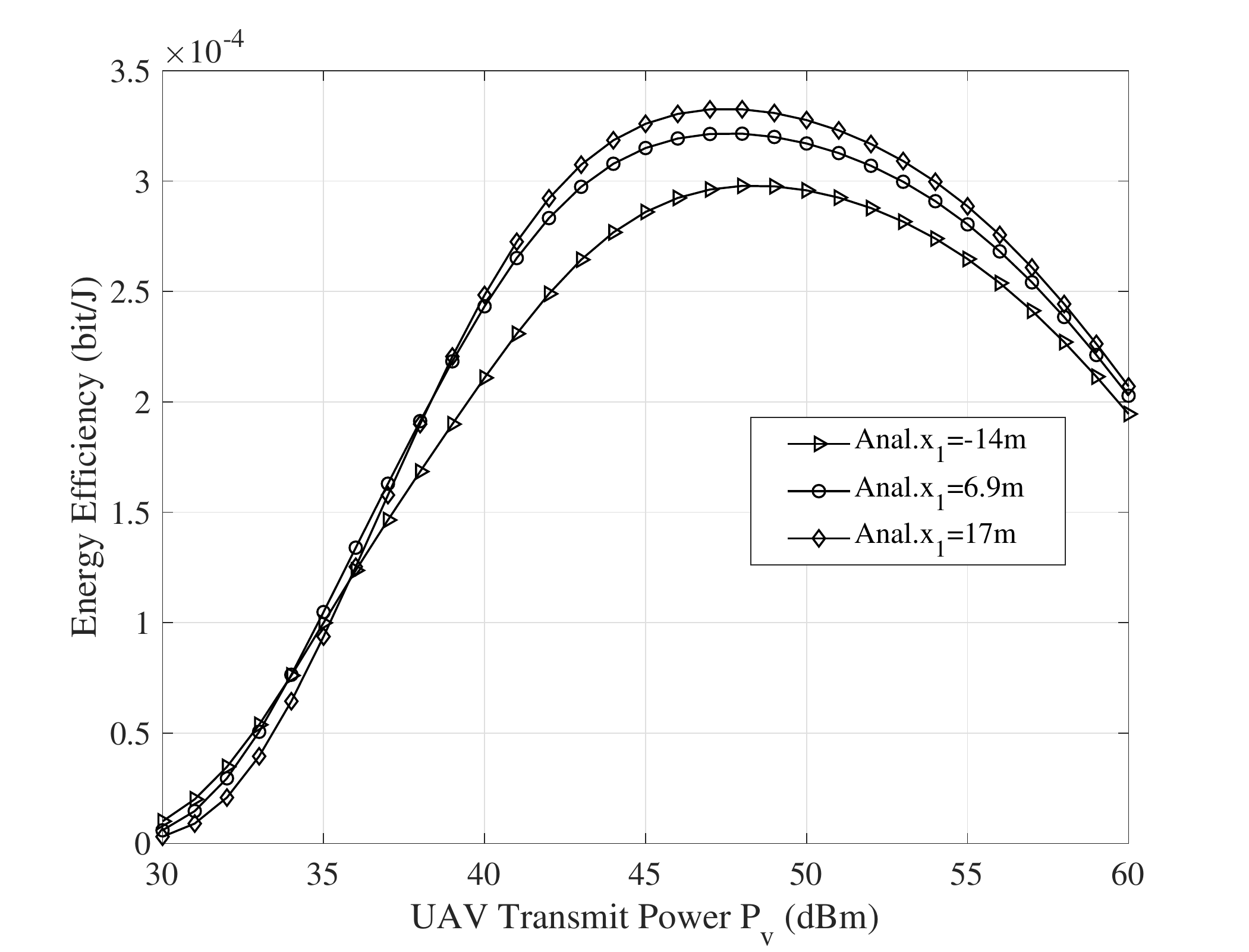}}
\caption{Energy efficiency optimization with different variables}
\label{EEwithInCons}
\end{figure}
Fig. \ref{EEwithInCons} plots the energy efficiency of the UAV versus (v.s.) its data collection locations and its transmit power. We can observe from Fig. \ref{EEwithInCons}(a) that higher UAV transmit power leads to closer UAV data collection location to the BS, and lower flight energy consumed by the UAV, contributing to enhanced energy efficiency. In addition, higher transmit power leads to lower average information outage probability, which also contributes to improved energy efficiency. Meanwhile, we can observe from Fig. \ref{EEwithInCons}(b) that there is an optimal transmit power leading to the maximum energy efficiency. However, the optimal transmit power on board should be feasible.

\section{Conclusion}
\label{section_conclusion}

In this paper, a UAV-assisted backscatter communications via TDMA was studied, where a UAV acts as a data collector from multiple terrestrial backscattering tags. We derived the closed-form expression of the system average outage probability, and optimized the UAV data collection location under the energy constraint in order to achieve maximum energy efficiency. Simulation results verified our derived analysis and illustrated that there is a tradeoff between the UAV data collection location and the energy efficiency.

%The extension of this work under more practical constraints with UAV swarms can be considered in future works.
% or


\appendices
\section{Energy Outage Probability}
\label{Proof_Penm}

\textit{Proof}\quad The energy outage probability can be formulated as  
\begin{equation}
\label{Penm11}
	\begin{aligned}
{P_{\textsc{e},\textup{m}}} & = \Pr ( {{\varepsilon _m} < \frac{{{T_\textsc{b}}}}{M}{P_\textsc{c}}} )= \Pr ( {{{\left| {{g_{\textsc{v}{\textsc{u}_m}}}} \right|}^2} < \frac{{{P_\textsc{c}}}}{{(m - {\eta _\textsc{r}}){\eta _\textsc{c}}{P_\textsc{v}}}}} ) \\
& = {p_{\textsc{v}{\textsc{u}_m},\textsc{LoS}}} \cdot \Pr ( {{{\left| {{g_{\textsc{v}{\textsc{u}_m},\textsc{LoS}}}} \right|}^2} < \frac{{{P_\textsc{c}}}}{{\omega(m - {\eta _\textsc{r}}){\eta _\textsc{c}}{P_\textsc{v}}}}} )\\
+& {p_{\textsc{v}{\textsc{u}_m},\textsc{NLoS}}} \cdot \Pr ( {{{\left| {{g_{\textsc{v}{\textsc{u}_m},\textsc{NLoS}}}} \right|}^2} < \frac{{{P_\textsc{c}}}}{{\omega(m - {\eta _\textsc{r}}){\eta _\textsc{c}}{P_\textsc{v}}}}} ).
	\end{aligned}
\end{equation}
Here in Eq. (\ref{Penm11}), $\Pr ( {{{\left| {{g_{\textsc{v}{\textsc{u}_m},\textsc{LoS}}}} \right|}^2} < \frac{{{P_\textsc{c}}}}{{\omega(m - {\eta _\textsc{r}}){\eta _\textsc{c}}{P_\textsc{v}}}}} )$ and $\Pr ( {{{\left| {{g_{\textsc{v}{\textsc{u}_m},\textsc{NLoS}}}} \right|}^2} < \frac{{{P_\textsc{c}}}}{{\omega(m - {\eta _\textsc{r}}){\eta _\textsc{c}}{P_\textsc{v}}}}} )$ can be both calculated by the CDF corresponding to the PDF in Eq. (\ref{cdfgain}), respectively. This completes the proof of Eq. (\ref{Penm}).$\hfill\blacksquare$

\section{Proof of ${F_{{\gamma _{{\textsc{u}_m}\textsc{v}}}}}\left( x \right)$ and ${F_{{\gamma _\textsc{vb}}}}\left( x \right)$}
\label{Proof_FgammaUmV}

Note that $\gamma_{\textsc{u}_m\textsc{v}}$ is given in Eq. (\ref{gammaUmV}), thus the CDF of $\gamma_{\textsc{u}_m\textsc{v}}$ can be derived as 
\begin{equation}
\label{FgammaUmV}
\begin{aligned}
&{F_{{\gamma _{{\textsc{u}_m}\textsc{v}}}}}( x ) = \Pr ( {\frac{{{\eta _{\textsc{r}}}{P_\textsc{v}}{{\left| {{g_{\textsc{v}{\textsc{u}_m}}}} \right|}^2}{{\left| {{{g'}_{\textsc{v}{\textsc{u}_m}}}} \right|}^2}}}{{{{\left| {{g_{\textsc{v}{\textsc{u}_m}}}} \right|}^2}\sigma _{{\textsc{u}_m}}^2 + \sigma _\textsc{v}^2}} < x} )\\
&=\int_0^\infty  {\Pr ( {{{\left| {{{g'}_{\textsc{v}{\textsc{u}_m}}}} \right|}^2} < \frac{{x\left( {y\sigma _{{\textsc{u}_m}}^2 + \sigma _\textsc{v}^2} \right)}}{{{\eta _{\textsc{r}}}{P_\textsc{v}}y}}} )\cdot{f_{{{\left| {{g_{\textsc{v}{\textsc{u}_m}}}} \right|}^2}}}( y )\textup{d}y} .
\end{aligned}		
\end{equation}
where  ${\Pr ( {{{\left| {{{g'}_{\textsc{v}{\textsc{u}_m}}}} \right|}^2} < \frac{{x\left( {y\sigma _{{\textsc{u}_m}}^2 + \sigma _\textsc{v}^2} \right)}}{{{\eta _{\textsc{r}}}{P_\textsc{v}}y}}} )}$ can be calculated similarly with Eq. (\ref{Penm11}), ${f_{{{\left| {{g_{\textsc{v}{\textsc{u}_m}}}} \right|}^2}}}\left( y \right)$ is given in Eq. (\ref{cdfgain}).
%\begin{equation}
%\begin{aligned}\nonumber
%&{\Pr \left( {{{\left| {{{g'}_{\textsc{v}{\textsc{u}_m}}}} \right|}^2} < \frac{{x\left( {y\sigma _{{\textsc{u}_m}}^2 + \sigma _\textsc{v}^2} \right)}}{{{\eta _{\textsc{r}}}{P_\textsc{v}}y}}} \right)} \\
%& ={p_{\textsc{v}{\textsc{u}_m},{\textsc{LoS}}}}\cdot \Gamma \left( k_{\textsc{vu}_m,\textsc{LoS}},\frac{{x\left( {y\sigma _{{\textsc{u}_m}}^2 + \sigma _\textsc{v}^2} \right)d_{\textsc{v}{\textsc{u}_m}}^2}}{{{\eta _{\textsc{r}}}{P_\textsc{v}}y{\beta _0}}} \right)\\
%& + {p_{\textsc{v}{\textsc{u}_m},{\textsc{NLoS}}}}\cdot \Gamma \left( k_{\textsc{vu}_m,\textsc{NLoS}},\frac{{x\left( {y\sigma _{{\textsc{u}_m}}^2 + \sigma _\textsc{v}^2} \right)d_{\textsc{v}{\textsc{u}_m}}^2}}{{{\eta _{\textsc{R}}}{P_\textsc{v}}\cdot y\cdot {\eta _{\textsc{v}{\textsc{u}_m}}}{\beta _0}}}\right).
%\end{aligned}		
%\end{equation}

Similarly, the CDF of $\gamma _{\textsc{vb}}$ can be derived as
\begin{equation}
\label{FgammaVB}
\begin{aligned}
{F_{{\gamma _\textsc{vb}}}}( x ) &= \Pr ( {{\gamma _\textsc{vb}} < x} )= {p_{\textsc{vb},{\textsc{LoS}}}}\cdot \Gamma ( {{k_{\textsc{vb},{\textsc{LoS}}}},\frac{{x\sigma _\textsc{b}^2d_\textsc{vb}^2\cdot {k_{\textsc{vb},{\textsc{LoS}}}}}}{{{P_\textsc{v}}{\beta _0}}}} )\\
 &+ {p_{\textsc{vb},{\textsc{NLoS}}}}\cdot \Gamma ( {{k_{\textsc{vb},{\textsc{NLoS}}}},\frac{{x\sigma _\textsc{b}^2d_\textsc{vb}^2\cdot {k_{\textsc{vb},{\textsc{NLoS}}}}}}{{{P_\textsc{v}}{\eta _\textsc{vb}}{\beta _0}}}}).
\end{aligned}		
\end{equation}
This completes the proof. $\hfill\blacksquare$

%% you can choose not to have a title for an appendix
%% if you want by leaving the argument blank
%\section{}

%\section*{Acknowledgment}
%
%
%The authors would like to thank China Scholarship Council (CSC) for funding the visiting study of Shengzhi Yang from Oct 2018 to Oct 2019 to King's College London (KCL), London, UK.

\ifCLASSOPTIONcaptionsoff
  \newpage
\fi




\bibliographystyle{IEEEtran}
\bibliography{ref}



%\end{IEEEbiography}
%% insert where needed to balance the two columns on the last page with
%%\newpage


%\vfill


\end{document}